\documentclass[10pt, conference, compsocconf]{IEEEtran}
  \usepackage{graphics}
  \usepackage{setspace}
  \usepackage{graphicx}
  \usepackage{epstopdf}
  \usepackage{color}

  \usepackage{float}
  \usepackage{wrapfig}
  \usepackage{color}

  \let\labelindent\relax
  \usepackage{enumitem}

  \usepackage{amssymb}
  \usepackage{latexsym}
  \usepackage{amsfonts}
  \usepackage{amsthm}
  \usepackage{amsmath}
  \usepackage{graphics}
  \usepackage{setspace}
  \usepackage{graphicx}
  \usepackage{epstopdf}
  \usepackage{color}
  \usepackage{float}
  \usepackage{wrapfig}
  \usepackage{color}
  \usepackage{rotating}
  \usepackage{array} 
  \usepackage{dblfloatfix}
  \usepackage[hyphens]{url}
  \usepackage[table]{xcolor}
  \usepackage{multirow}

\ifCLASSINFOpdf
\else
\fi
\begin{document}
%
\title{CA4IOT: Context Awareness for Internet of Things \vspace{-0.4cm}}




%

\author{\IEEEauthorblockN{Charith Perera\IEEEauthorrefmark{1}\IEEEauthorrefmark{2},
Arkady Zaslavsky\IEEEauthorrefmark{2},
Peter Christen\IEEEauthorrefmark{1} and
Dimitrios Georgakopoulos\IEEEauthorrefmark{2}}
\IEEEauthorblockA{\IEEEauthorrefmark{1}Research School of Computer Science, The Australian National University,Canberra, ACT 0200, Australia}
\IEEEauthorblockA{\IEEEauthorrefmark{2}CSIRO ICT Center, Canberra, ACT 2601, Australia}}

%


\maketitle

\begin{abstract}
Internet of Things (IoT) will connect billions of sensors deployed around the world together. This will create an ideal opportunity to build a sensing-as-a-service platform. Due to large number of sensor deployments, there would be number of sensors that can be used to sense and collect similar information. Further, due to advances in sensor hardware technology, new methods and measurements will be introduced continuously. In the IoT paradigm, selecting the most appropriate sensors which can provide relevant sensor data to address the problems at hand among billions of possibilities would be a challenge for both technical and non-technical users. In this paper, we propose the Context Awareness for Internet of Things (CA4IOT) architecture to help users by automating the task of selecting the sensors according to the problems/tasks at hand. We focus on automated configuration of filtering, fusion and reasoning mechanisms that can be applied to the collected sensor data streams using selected sensors. Our objective is to allow the users to submit their problems, so our proposed architecture understands them and produces more comprehensive and meaningful information than the raw sensor data streams generated by individual sensors.

\end{abstract}


%
\IEEEpeerreviewmaketitle

\section{Introduction}
\label{sec:Introduction}

\begin{figure}[b]
 \centering
 \vspace{-0.43cm}
 \includegraphics[scale=.40]{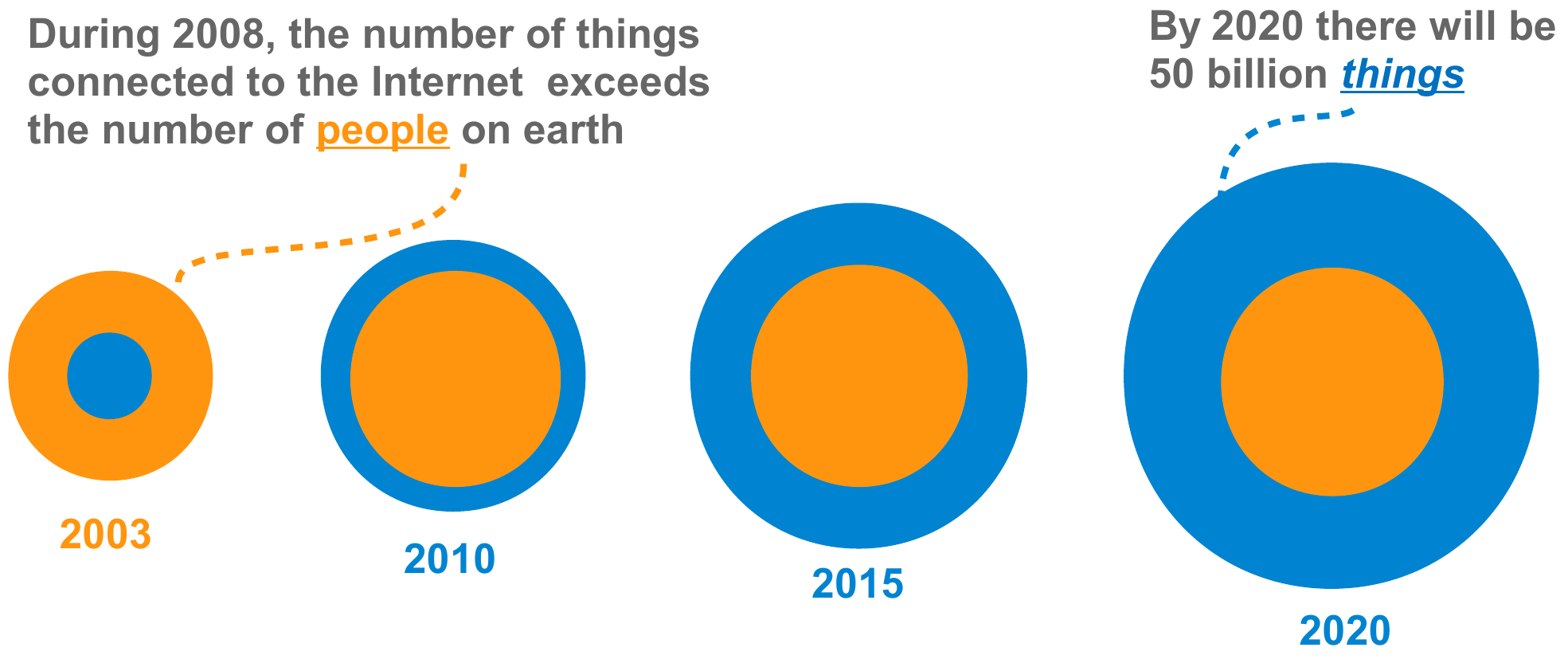}
\vspace{-0.43cm}	
 \caption{Growing number of things connected to the Internet \cite{P574}}
 \label{Growing number of things connected to the Internet}	
\vspace{-0.23cm}	
\end{figure}

The Internet of Things (IoT) is the next phase of the evolution of the Internet. The internet has passed several phases since it was invented in the early 1980s. The Internet expanded from few computers communicating with each other to billions of computational nodes to billions of mobile phones over the time. Now it is moving towards a phase where all objects around us will be connected to the Internet and will be able to communicate with each other. The European Commission has predicted that by 2020 there will be 50 to 100 billion devices connected to the Internet  \cite{P029}. As depicted in Figure \ref{Growing number of things connected to the Internet}, the number of things connected to the Internet exceeds the number of people on Earth in 2008.

The things that we expect to connect to the Internet will comprise sensors, actuators and processing capability that will make themselves intelligent. In that perspective, connecting sensors and actuators together over a network is not totally new to the field of computer science and engineering. Sensor networks \cite{P009} have been used and researched over many decades. However in early days, the focuses were on building specialised applications such as detecting a wild fire in a forest. Further research efforts were focused on low-level operations such as energy optimisation, routing, mobility and reliability.  In contrast, the IoT is more focused on high-level tasks such as collecting, storing, querying, and understanding sensor data. However, sensor networks are the backbone of the IoT.

When large numbers of sensors are deployed and start collecting data, traditional application based approaches becomes infeasible. Therefore, significant amount of middleware solutions have been introduced by researchers. An evaluation and comparison of a subset of available middleware solutions that focused on sensor networks, pervasive/ubiquitous computing, and the IoT are presented in \cite{P035}, \cite{P417}, \cite{P118}, \cite{P291}. Each middleware solution focuses on different aspects in IoT such as device management, interoperability, platform portability, context-awareness, security and privacy and many more. Even though, some solutions address multiple aspects, an ideal middleware solution that addresses all the aspects required by the IoT is yet to be designed.

Our objective is not to introduce such an ideal middleware solution. Our goal is to design an solution to help users to automating the task of selecting the sensors according to the problems/tasks at hand. Further Explanations are provided in Section II. Our proposed approach, called CA4IOT, can be adopted into any IoT middleware solution.

The paper is organised in sections as follows. Section \ref{sec:Problem Definition and Motivation} defines the problem and the motivations to address it. Section \ref{sec:Functional Requirements} identifies the functional requirements that need to be addressed in order to solve the problem. In Section \ref{sec:Layered Architecture}, we overview the CA4IOT layered architecture. Section \ref{sec:The CA4IOT Architecture} presents a detailed explanation of CA4IOT architecture in component level. In Section \ref{sec:Use Case} we introduce a use case to further explain the execution process of CA4IOT architecture with justifications in step by step. Finally, Section \ref{sec:Conclusions and Future Work} presents some concluding remarks and future work.

\section{Problem Definition and Motivation}
\label{sec:Problem Definition and Motivation}

\begin{figure}[h]
 \centering
 \includegraphics[scale=.45]{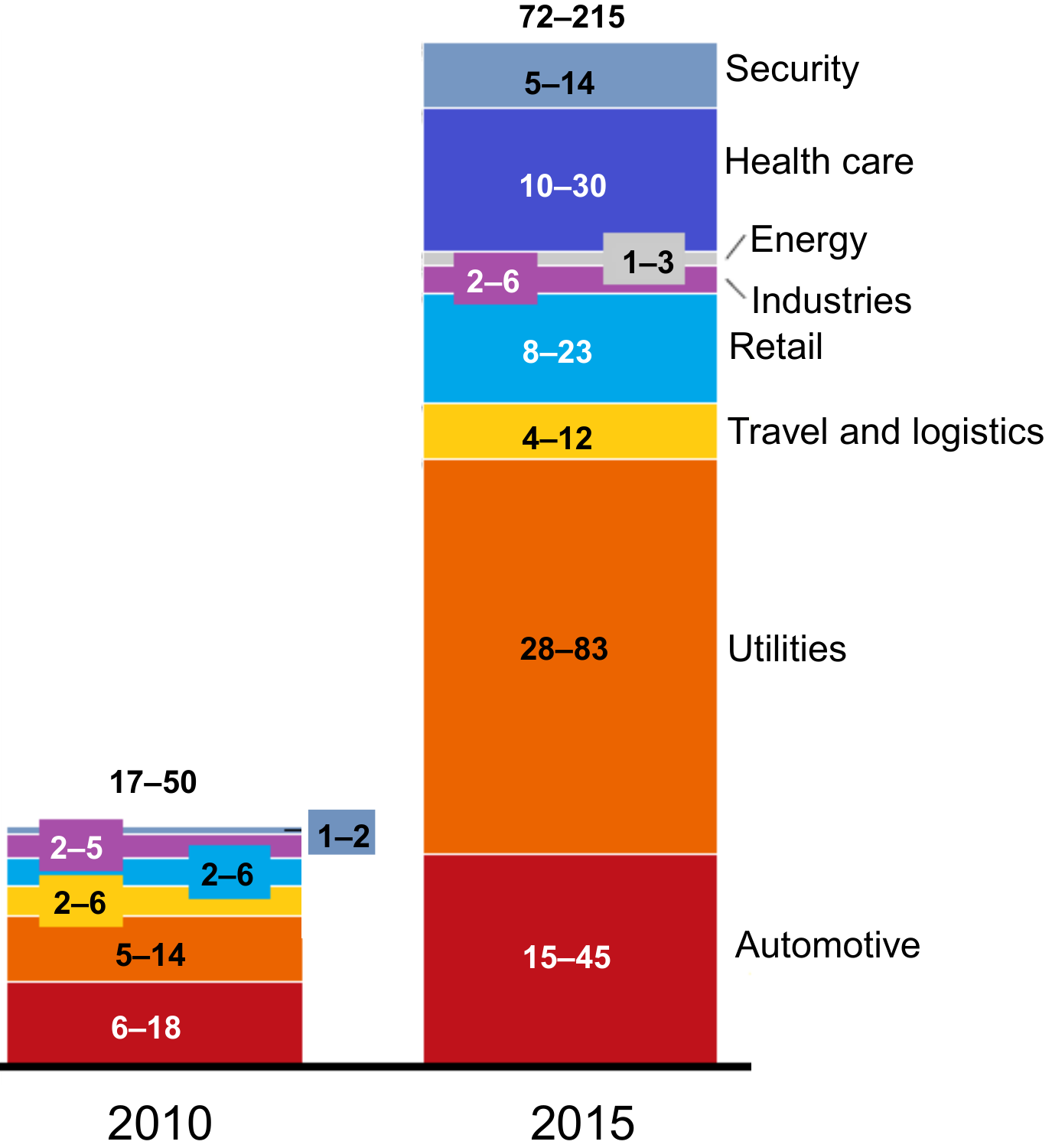}
\vspace{-0.23cm}	
 \caption{Estimated numbers of things (nodes) connected to the Internet based on sectors are presented in millions \cite{P504}}
 \label{Number of Sensor nodes and Industries}	
\vspace{-0.23cm}	
\end{figure}

The IoT envisions an era where billions of things are connected to the Internet.  In Figure \ref{Number of Sensor nodes and Industries}, the predicted growth of things connected to the Internet is presented based on sectors. Utilities, automotive, healthcare and retail industries will contribute to the growth significantly.

Let's define the problem in detail. In order to deal with growing number of sensors, significant numbers of middleware solutions are proposed. These middleware solutions such as GSN \cite{P022}, \cite{P227}, provide high-level data processing capabilities such as querying, filtering, and fusing. However, in the existing state of the art middleware solutions users have to select the sensors they want. For example, let's consider an environmental scientist who is studying and analysing environmental pollution. He wants to measure environmental pollution in Canberra, Australia. In the current approaches, he should know what sensors measure environmental pollution, how many relevant sensors are connected to the middleware solution he is using, the specific locations in term of GPS coordinates, and so on.

\begin{figure}[h]
 \centering
 \includegraphics[scale=.45]{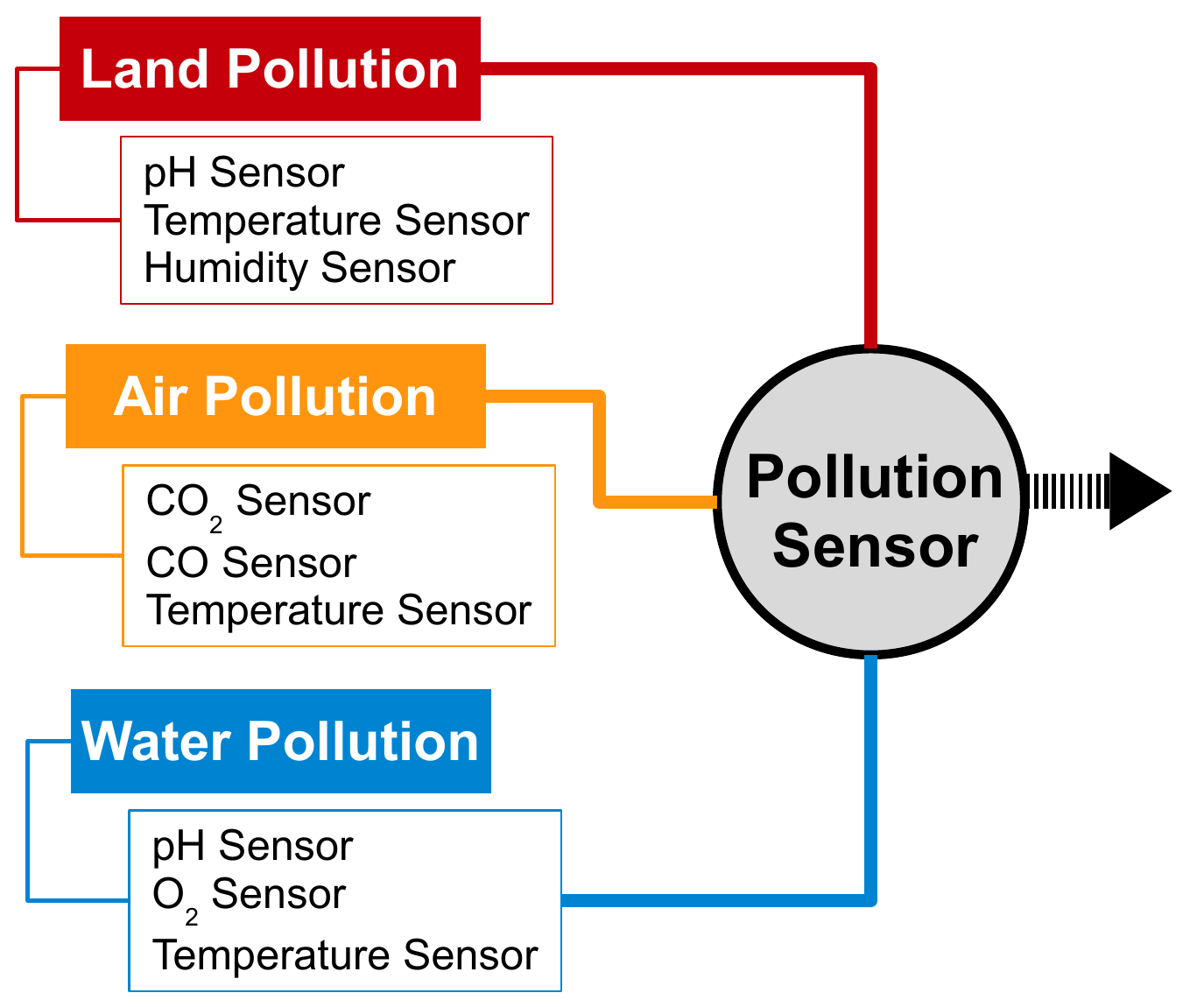}
 \caption{Gap between high-level user requirements and low-level sensors}
 \label{Sensor_Gap}	
\vspace{-0.33cm}	
\end{figure}

This would not be an issue if there are only few hundreds of sensors. However, we are moving towards an era where billions of sensors would be available to use through middleware solutions. In this situation, manually selecting the relevant sensors is not feasible. Users such as environmental scientists are non-technical personals who do not have extensive knowledge in computer science. They are only interested in acquiring relevant data so they can use the data to build models, simulations, understand and solve their problems. Therefore, there is a clear gap between what the user wants and what is available. We can further explain the problem using Figure \ref{Sensor_Gap}. Based on the scenario we introduced previously, there is no single sensor that is capable of measuring environmental pollution. For example, environmental pollution can be simply attributed in to three sub categories: land pollution, air pollution, and water pollution. Each category can be measured by large number of sensors. Three example sensors are depicted in Figure \ref{Sensor_Gap}. This illustration provides a way to understand how the manually selection of sensors could be extremely complex. \textit{``How to reduce the complexity of selecting appropriate sensors by understanding the user requirements /problems?"} is the problem we have addressed in this paper. Ideally, IoT middleware solutions should allow the users to express what they want and provide the relevant sensor data back to the users quickly without asking the users to manually selecting sensors which relevant to their requirements. In this paper, we propose an architectural approach to automate the configuration of filtering, fusing and reasoning sensors according to the user's requirements. Specifically, when a user requests environmental pollution measurements in Canberra, our approach combines all relevant sensor data together and provides to the user as a single data stream so the user can feed them to their own system to extract further information on environmental pollution.

There are several other motivations to address this problem. In the recent past, cloud computing \cite{P498} and everything-as-a-service (XaaS) have gained significant attention. Therefore, sensing-as-a-service \cite{ZMP001} envisions to provide sensing capabilities as a service over cloud technologies to the consumers. Such a model will be benefited by our approach, because the consumers are non-technical. In addition, even though there are significant amount of IoT and sensor network middleware solutions proposed, the problem we focus here is largely unaddressed and unattended. In addition, our solution can be used as a service so other innovative application can be built on top.

\section{Functional Requirements}
\label{sec:Functional Requirements}

In order to design the architecture for CA4IOT, we evaluate the problem in depth to identify the functional requirements of the proposed solution. These functionalities are reflected in the CA4IOT architecture discussed in details in Section \ref{sec:The CA4IOT Architecture}. Some of the ideas presented in this section will get clearer when we explain the execution process using an example scenario in Section \ref{sec:Use Case}.

\textbf{Ability to connect sensors to the IoT middleware easily:} This is an important functionality where billions of sensors are expected to be connected to IoT middleware solutions. Due to the scale, it is not feasible to connect sensors manually by technical people. We demonstrated how the connecting sensors can be automated in \cite{ZMP002}. In addition,  IEEE 1451 \cite{P257} and SensorML \cite{P083} allow to make the automation process more sophisticated.

\textbf{Ability to understand and maintain context information (what, when, who, how, why) about sensors:} Context information about sensors needs to be acquired and stored with appropriate annotations which make it easy to retrieve them later. Up-to-date information such as sensor capabilities, location, sampling rate, nearby sensors, battery life, etc. need to be maintained. This knowledge is required to select appropriate sensors based on the users' request.

\textbf{Ability to understand the user requirement / request / problem:} CA4IOT needs to reason and understand the user request. For example, as explained in Figure \ref{Sensor_Gap}, CA4IOT should be able to understand the relationship between environmental pollution and low-level sensors such as temperature sensors and pH sensors. This can be achieved by maintaining knowledge about application domains using knowledge bases. In addition, these knowledge bases should be able to extend easily by plugin additional knowledge bases which contain knowledge on different application domains when necessary.

\textbf{Ability to fill the gap between high-level user requirements and low-level sensors capabilities:} Reasoning (e.g. semantic or statistical) is essential to understand the relationship between high-level user requirements and low-level sensor capabilities as explained in Figure \ref{Sensor_Gap}. Further reasoning is required to identify relevant context information based on given sensor reading and also to generate new knowledge (e.g. read GPS location coordinates of two sensors and decide they are nearby).

\textbf{Ability to extract high-level context information using low-level raw sensor data:} There are many operations that can be applied to the sensor data. An operation could be as simple as averaging or as complex as combining multiple sensor readings and calculate a single reading or generating missing values by evaluating historic sensor data. Mostly, data fusion operations are used to generate new context knowledge.

\textbf{Ability to manage users:} This is about acquiring, reasoning and storing user information. When users make requests, CA4IOT needs to keep track of them in term of the their requirements, output format required, additional constraints such as sampling rate, data volume, and so on. Further, CA4IOT needs to provide a mechanism to interact with the users which will allow to define their requirement easily.

\begin{figure}[h]
 \centering
 \includegraphics[scale=.60]{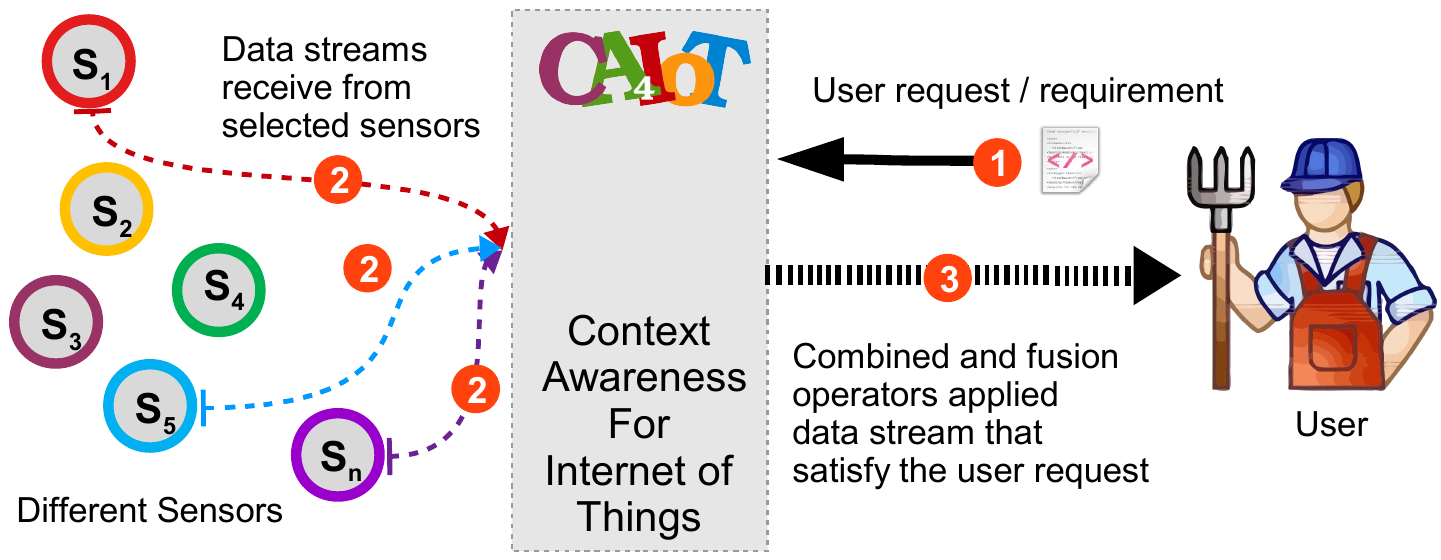}
\vspace{-0.33cm}	
 \caption{Functional objective of CA4IOT}
 \label{Functional objective of CA4IOT}	
\vspace{-0.43cm}	
\end{figure}

These functionalities allow us to achieve our main objective as depicted in Figure \ref{Functional objective of CA4IOT} in three steps. In Step 1, user provides his requirement to CA4IOT. CA4IOT understands the user problem and selects the appropriate sensors as shown in step 2. In step 3, CA4IOT combines the sensor data retrieved from selected sensors as a single data stream and sends them to the user. The next section provides an overview on layered architecture of CA4IOT.

\section{Layered Architecture}
\label{sec:Layered Architecture}

\begin{figure}[h]
 \centering
 \includegraphics[scale=.50]{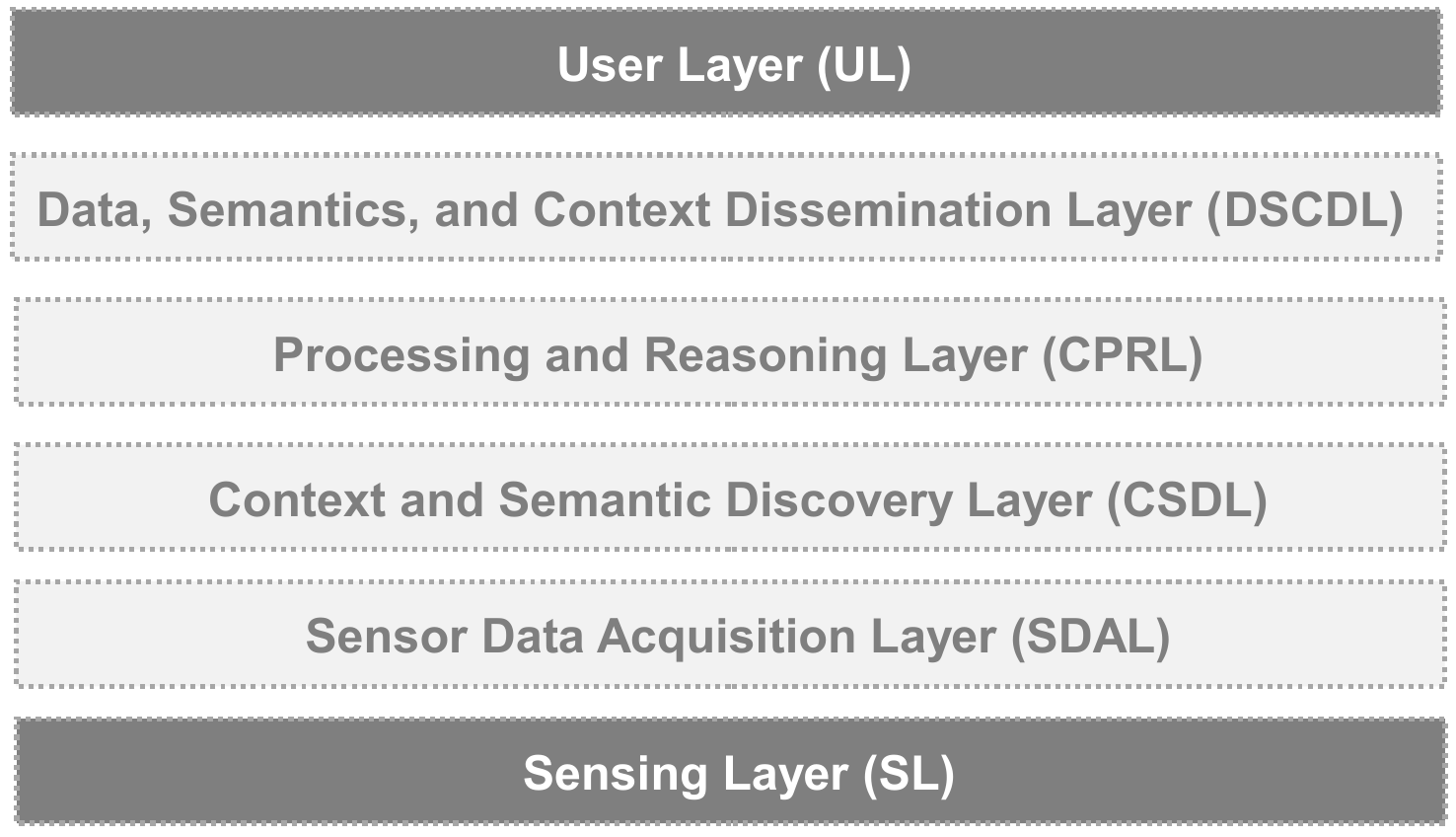}
\vspace{-0.23cm}	
 \caption{CA4IOT architecture consists  of four internal layers and two external layers}
 \label{CA4IOT architecture}	
\vspace{-0.43cm}	
\end{figure}

This section provides a high-level overview of the CA4IOT architecture based on layers. As we mentioned earlier, CA4IOT is not focused on providing a complete software solution that addresses all the requirements needed in IoT paradigm. Instead, we focus on a single problem as discussed in Section \ref{sec:Problem Definition and Motivation}. As a result, layered architecture depicted in Figure \ref{CA4IOT architecture} intended only to address that specific problem. We recommend not to employ CA4IOT as a standalone middleware, but to combine the architecture, models, and techniques into existing IoT middleware solutions which intend to fulfill the demands in IoT paradigm.

The CA4IOT architecture consists of four layers: \textit{Data, Semantics, and Context Dissemination Layer (DSCDL), Processing and Reasoning Layer (CPRL), Context and Semantic Discovery Layer (CSDL),} and \textit{Sensor Data Acquisition Layer (SDAL)}.

However, there are two other layers (User layer and sensing layer) that interact with CA4IOT. They are not part of CA4IOT. However, they are essential for a successful interaction and execution. Each layer is designed to perform a specific action. Most of the IoT, sensor network, and context management middleware solutions always have similar layers with similar names. We have named each layer based on its responsibility. The following describes each layer briefly. Here, we introduce the components that belong to each layer though we discuss the components in detail in Section \ref{sec:The CA4IOT Architecture}.

\textbf{User Layer (UL):} This is the layer that represents the users and it is not a core layer in CA4IOT. Users can be human users, applications, or services. User Oriented Front End (UOFE) is a part of this layer and therefore, it is not a core component in CA4IOT architecture.

\textbf{Data, Semantics, and Context Dissemination Layer (DSCDL):} This layer is responsible to manage users. The components belong to this layer are data dispatcher, request manager, and publish/subscribe. 

\textbf{Processing and Reasoning Layer (CPRL):} This is the most important layer in CA4IOT. It is responsible for data processing, reasoning, fusing, knowledge generating and storing. The components belong to this layer are context registry, context knowledgebase, reasoning engine, context and semantic discoverer generator, primary context processing, secondary context processor, context provider registry, data fusion operator, and data fusion repository.

\textbf{Context and Semantic Discovery Layer (CSDL):} This layer is responsible for managing context and semantic discoverers which includes generating, configuring, and storing. The components belong to this layer are context and semantic discoverers, context and semantic discoverer generator, and context and semantic discoverers repository.

\textbf{Sensor Data Acquisition Layer (SDAL):} This layer is responsible for acquiring data. This layer appears in most the IoT, sensor network, and context management middleware solutions with different terminologies such as wrappers, gateways, handlers, proxies, mediators, etc. This layer communicates with hardware and software sensors and retrieves sensor data into CA4IOT. The components that belong to this layer are sensor wrappers, wrapper repository, wrapper generator, sensor device definition (SDD) local repository, and SDD cloud repository.

\textbf{Sensing Layer (SL):} This layer represents all software and hardware (physical and virtual) sensors. Further, this layer is not a part of core CA4IOT architecture.

\section{The CA4IOT Architecture}
\label{sec:The CA4IOT Architecture}

The CA4IOT architecture consists of several components. In Section \ref{sec:Layered Architecture}, we introduced all the components under each layer. In this section, we discuss each component including their primary responsibilities. Figure \ref{CA4IOT component-level architecture} illustrates the component-level architecture of CA4IOT. We have labelled each component (e.g. C1, C2). However, numbering does not reflect the execution or interaction order. Further, there are some other elements that do not belong to CA4IOT architecture but essential to be explained as they are evolve in execution process. These elements are also labelled (e.g. E1, E2).

The order of the components we explain does not necessarily correspond to the order of execution. We explain the execution process with an example scenario in Section \ref{sec:Use Case}. We specify the possibility of multiple components using the “(s)” notation. For example, multiple sensor wrappers are defined as sensor wrapper(s).

\begin{itemize}[noitemsep,nolistsep,leftmargin=1em]

\vspace{-0.33cm}
 \item \textbf{User(s) (E1):} User in CA4IOT can be a human user, application or a service. Users interact with User Oriented Front End(s) (UOFE) to express their requirements.

\item \textbf{User Oriented Front End(s) (UOFE) (E2):} This element is not a part of CA4IOT architecture. UOFE can be a graphical user interface (GUI), a web service, a natural user interface, or any other mechanism that allows the users to express their requirement in high-level (in abstract). UOFE element generates a XML file, \textit{request.xml}, based on the user requirements according to the schema definition provided by CA4IOT. We use XML to decouple the mechanism of users expressing their requirements from the internal CA4IOT execution process. Therefore, it significantly increases the flexibility and creativity which allows developers to develop more sophisticated mechanisms for the users to express their requirements. From CA4IOT perspective, we expect only a XML file that complies to our request definition. The mechanism that used to create the \textit{request.xml} does not make any impact on CA4IOT.

\item \textbf{Sensor ($S_{n}$) (E3):} It can be either physical or virtual. For example, a physical sensor can be a temperature sensor. In contrast, a virtual sensor can be a web service hosted by some organisation that provides information (e.g. weather information or business contact information).

\item \textbf{Data Output Mechanisms/Formats (E4):} There are many output mechanisms and data formats. For example, data can be archived in a cloud repository. In addition, data can be inserted into a multimodal interface and visualisation program. Sensor data can also be produced as open linked data. Some of the popular data formats are XML, CVS, and JSON.

\begin{figure*}[t]
 \centering
 \includegraphics[scale=.68]{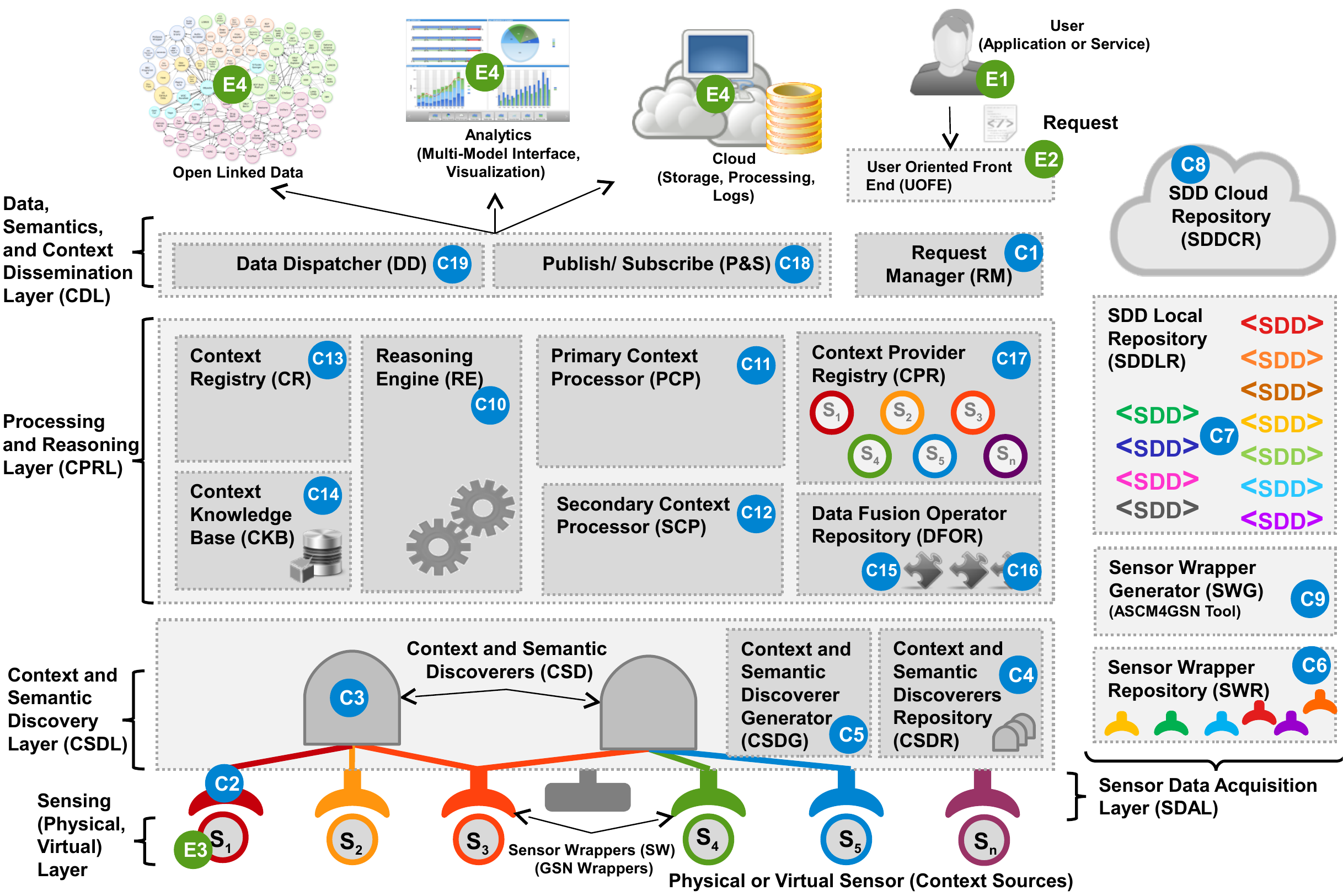}
\vspace{-0.23cm}	
 \caption{CA4IOT component-level architecture}
 \label{CA4IOT component-level architecture}	
\vspace{-0.53cm}	
\end{figure*}

\item \textbf{Request Manager (RM) (C1):} It is responsible for handling user requests. Users submit their requirements using an XML called \textit{request.xml}. First, RM performs a validation to make sure that the received request complies with the specification. Then, RM extracts the information and converts them into number of objects which can be passed among different component programmatically. User details are sent to the publish/subscribe component to be stored for later retrieval.

\item \textbf{Sensor Wrapper(s) (SW) (C2):} It is responsible to communicate with physical and virtual sensors. Specifically, they are used to retrieve data from sensors. For example, when connecting to a physical sensor to CA4IOT, manufacturer released sensor specific hardware APIs need to be used within this component. These components can be generated automatically using \textit{sensor device definition} (SDD) files as explained in \cite{ZMP002} or can be manually developed by programmers. In order to communicate with a specific sensor, CA4IOT should have a corresponding wrapper attached it. For example, in order to communicate with SunSPOT \cite{P382} sensor, CA4IOT should have a SunSPOT wrapper that is capable of communicate with SunSPOT sensor. However, in some circumstances single wrapper can be used to communicate with different types of devices as mentioned in \cite{ZMP001} if the devices follow same communication protocols and  hardware APIs.

\item \textbf{Sensor Wrapper Repository (SWR) (C6):} It holds all the sensor wrappers that have been created before. When CA4IOT wants to retrieve sensor data from a senor, a request will be sent to SWR to check whether there is a corresponding wrapper available in the repository. If found, the wrapper will be assigned to the CSD. If not, the request will be forwarded to the SDD local repository.

\item \textbf{SDD Local Repository (SDDLR) (C7):} It is responsible to handle SDD files locally. When CA4IOT could not find a wrapper in SWR, CA4IOT will send a request to SDDLR to check whether there is a SDD file that can be used to generate the wrapper that is needed. If found, the SDD file will be sent to SWG to generate the wrapper based on the SDD.

\item \textbf{SDD Cloud Repository (SDDCR) (C8):} This is same as SDDLR but resides in the cloud. Developers around the world can submit the sensor wrappers to this repository as explained in \cite{ZMP002}. SDDLR can communicate with SDDCR to retrieve SDD files that are not available in SDDLR.

\item \textbf{Sensor Wrapper Generator (SWG) (C9):} It generates the sensor wrappers based on sensor device definition (SDD) files and send them to SWR.

\item \textbf{Context and Semantic Discoverer(s) (CSD) (C3):} These components are specifically custom build to satisfy user requirements. That means each CSD is responsible to satisfy one user request. Further, each CSD can communicate with multiple sensor wrappers to retrieve sensor data. CSD's main responsibility is to collect sensor data and bundle them together to satisfy the user requirement. CSD uses data fusion operators to transform and extract high-level information using raw sensor data as the user specified in the request. These components are system generated based on the reasoning output. After retrieving and applying data fusion operators, CSD starts sending data to the data dispatcher in order to be sent to the user.

\item \textbf{Context and Semantic Discoverers Repository (CSDR) (C4):} It holds all the CSDs created before. As we mentioned earlier, each CSD is custom built to address one user request. However, a CSD can be reused if another or same user makes exactly the same request. CSDR can search its repository to find whether there is a CSD that is created before that can satisfy a given user request.

\item \textbf{Context and Semantic Discoverers Generator (CSDG) (C5):} It is responsible for generating CSDs based on the specification given by reasoning engine.

\item \textbf{Reasoning Engine (RE) (C6):} It performs number of reasoning tasks using semantic and statistical reasoning techniques \cite{P185}. RE analyses the user problem and reason what context information is required to satisfy the user. Further, it handles the entire execution process of CA4IOT. It is the central component that monitors and makes decisions on the execution process.

\item \textbf{Primary Context Processor (PCP) (C11):} Once the RE identifies the context information that need to be collected in order to fulfill the user requirement, PCP identifies how to capture the required context data using existing sensors. PCP communicates with CPR and CKB to make the final decision considering many factors such as cost, availability, data quality, etc.

\item \textbf{Secondary Context Processor (SCP) (C12):} Secondary context is any piece of context data that can be computed using primary context data. SCP selects appropriate sensors from CPR and use data fusion operators from DFR to decide the best mechanism to capture secondary context information that is required by the user. CKB is used to reason the domain knowledge when required.

\item \textbf{Context Registry (CR) (C13):} It maintains a registry of possible context information that can be captured by using sensors. Its responsibility is to help the reasoning engine to extract context information that is required to fulfill the user requirement. While CKB acts as a source of domain knowledge, CR acts as a source of knowledge on sensors and related information.

\item \textbf{Context Knowledge Base (CKB) (C14):} It is responsible to store and reason domain knowledge that is required to understand the user requirements. Users are allowed to express their requirement in high-level. These high-level descriptions are domain specific. In order to understand the user requirement, CA4IOT needs to maintain domain knowledge (e.g. agriculture domain, smart home domain, etc.) This CKB supports plugin architecture so new domain knowledge can be added when necessary. Due to plugin architecture, CKB need to be defined according to the specification provided. Specification confirms smooth interoperability with existing knowledge.

\item \textbf{Data Fusion Repository (DFR) (C15):} It holds all the data fusion operators and provides functionalities such as searching and reasoning. New data fusion operators can be added to the repository as it supports plugin architecture. However, data fusion operators need to be designed using given specification which specify some essential data structures and functions that increase the interoperability of the data fusion operators. DFR maintains a comprehensive description of all DFOs in the repository as they helps to select the appropriate DFO that can deliver the expected results.

\item \textbf{Data Fusion Operator(s) (DFO) (C16):} Each data fusion operator is designed to accomplish one task. For example, missing GPS value generation operator is designed to generate missing value based on historic data and predictive algorithms. There can be multiple data fusion operators designed to accomplish the same task. The appropriate data fusion operator for each task is selected by RE and SCP depending on the user requirements.

\item \textbf{Context Provider Registry (CPR) (C17):} It is responsible to keep track of all the context providers (i.e. sensors). All the sensors that connect to CA4IOT are registered in CPR by providing all the information such as its capabilities and availability. This help to identify the available sensors to be used to satisfy the user requests when required. Once the RE identifies which context information is required, CPR is used to find the sensors that provide the matching context information (sensor data) that can be used to fulfill the user requirement.

\item \textbf{Publish/Subscribe (P/S) (C18):} Information about the users is stored here so they can be retrieved later once the sensor data is prepared to be delivered. User information stored in P/S helps to find out delivery requirements such as frequency, data format, etc.

\item \textbf{Data Dispatcher (DD) (C19):} It is responsible to deliver the sensor data produced by CSD to the user based on the user requirement such as frequency, format, etc. DD communicates with P/S to gather information about the user. In addition, all the CSDs produce its output in a CA4IOT standard output format. DD transform the data into user required format.

\end{itemize}

\section{Use Case}
\label{sec:Use Case}

\begin{figure}[h]
 \centering
 \includegraphics[scale=.40]{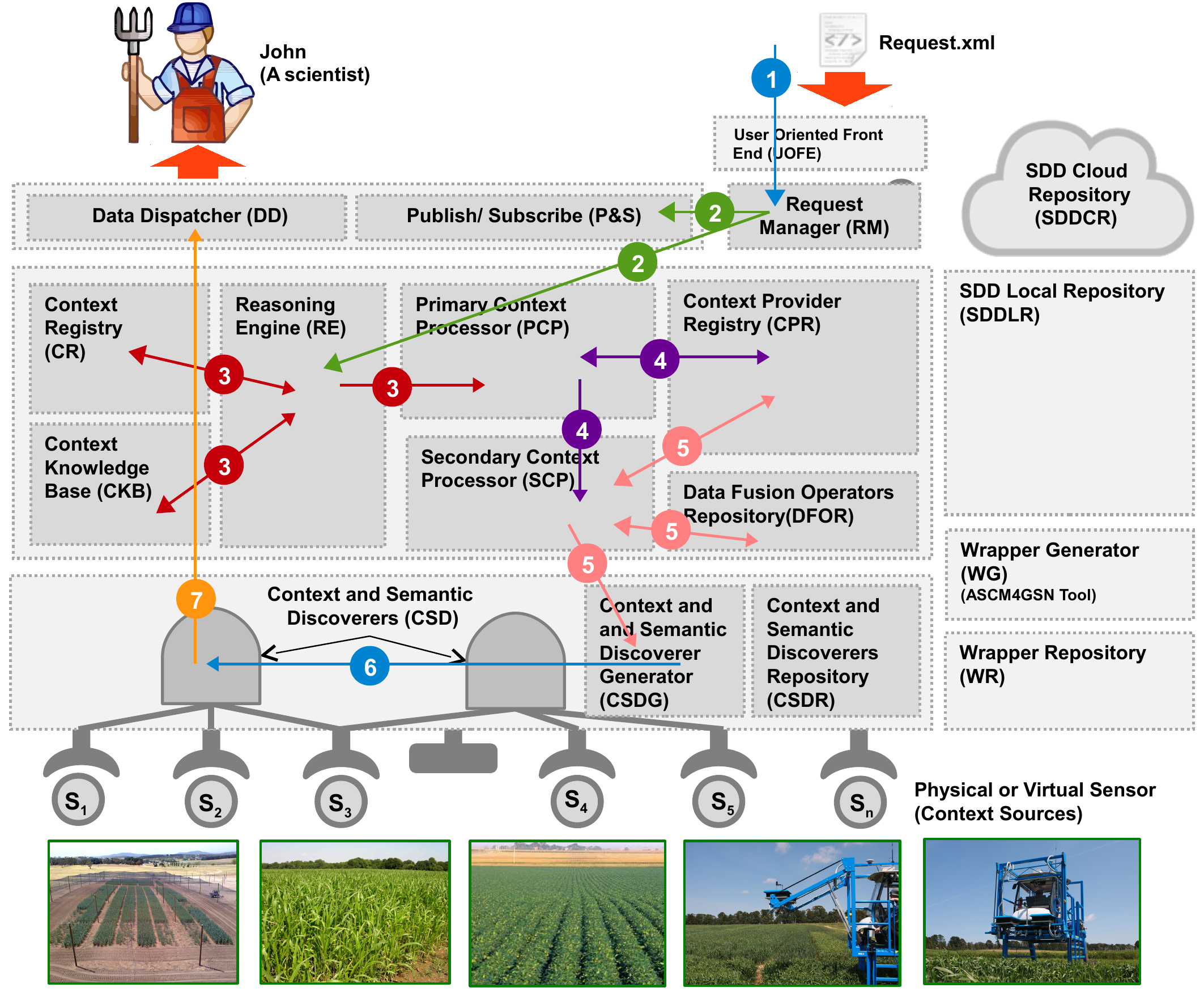}
\vspace{-0.53cm}	
 \caption{Execution and interaction process of CA4IOT}
 \label{Execution and interaction process of CA4IOT}	
\vspace{-0.43cm}	
\end{figure}

Previously, in Section \ref{sec:Layered Architecture} and \ref{sec:The CA4IOT Architecture}, we discussed CA4IOT architecture both in layered and component-level perspective. We discussed each component in isolated fashion with little focus on execution process and interaction among components. Still some of the fact may seem unclear without proper examples. Therefore, in this section, we provide a detailed scenario based on real-world smart agriculture domain with hypothetical facts. We explain each step from the beginning to the end in order as depicted in Figure \ref{Execution and interaction process of CA4IOT}. There are number of different execution processes that can be occurred in CA4IOT. However, in here, we focus only on explaining the scenario where user submits a request and CA4IOT provides the relevant data to the user. First, we explain the background information and then present the execution process and interaction details with the help of background information and hypothetical facts.

\subsection{Background Information}
\label{sec:UC:Background Information}

Every year, Australian grain breeders plant up to 1 million 10m$^{2}$ plots across the country to find the best high yielding varieties of wheat and barley. The plots are usually located in remote places often requiring more than four hours travel one-way to reach. The challenge is to monitor the crop performance and growing environment throughout the season and return the information in an easily accessible format. The challenge of crop growing and performance monitoring can be addressed by deploying sensors. Further, querying the collected sensor data is essential to understand what is happening in the field. In order to answer complex and sophisticated queries, significant amount of context data need to be stored with the raw sensor data. In addition, semantics also need to be attached to the raw sensor data.

Let's consider a scenario. John, a plant scientist, who is looking after a experimental crops growing facility, wants to know whether the crops are infected by \textit{Phytophtora} disease. \textit{Phytophtora} \cite{P452} is a fungal disease which can enter a field through a variety of sources. Humidity plays a major role in the development of \textit{Phytophtora}. Both temperature and whether or not the leaves are wet are also important indicators to monitor \textit{Phytophtora}. Based on the above real world information we created the following hypothetical facts. We assume that these are the only rules that make impact on detecting \textit{Phytophtora} disease (store in CKB).

\begin{itemize}
\footnotesize
 \item IF \textbf{airTemperature}  $\textless$ 12 AND \textbf{airHumidity} $\textless$ 25\% THEN \textbf{airStress} level = low 
 
 \item IF \textbf{airTemperature}  $\geq$ 12 AND \textbf{airHumidity} $\geq$ 25\% THEN \textbf{airStress} level = high  
 
 \item IF \textbf{airStress} = high  AND \textbf{leafWetness} $\textgreater$ 50 THEN \textit{Phytophtora disease} = infected ELSE = not-infected

\end{itemize}

\subsection{Execution Process}
\label{sec:UC:Execution Process}

\textbf{Step 1:} John, a plant scientist (i.e. user), expresses his request (\textit{Do any of the plots have infected by Phytophtora disease?}) using User Oriented Front End (UOFE). UOFE automatically generates a request, a \textit{request.xml} file, based on the user expression. Then, the request is sent to the Request Manager (RM). John also wants the data in the format of JSON in a frequency rate of 1 minute.

\textbf{Step 2:} RM validates the \textit{request.xml} and extracts the user information and send them to the Publish/Subscribe (P/S). Then, RM sends the user request to Reasoning Engine (RE).

\textbf{Step 3:} RE communicates with Context Knowledge Base (CKB) and Context Registry (CR) to identify the context information related to the users request. Based on the rules provided previously and the domain ontology support, RE identifies \textbf{\textit{airStress}} and the \textbf{\textit{leafWetness}} as the context information that is required to answer the John's request.
RE lists down all the context information required and sends the whole list to the Primary Context Processor (PCP).

\textbf{Step 4:} PCP communicates with Context Provider Registry (CPR) to detect the sensors which can provide the context information listed in the list provided by RE. PCP detects that \textbf{\textit{leafWetness}} can be directly acquired by a sensor as the primary context data. However, airStress cannot be identified directly from a sensor. Therefore, \textbf{\textit{airStress}} is sent to the Secondary Context Processor (SCP) for further processing. The context information that cannot be acquired directly using sensors are sent to SCP. 

\textbf{Step 5:} SCP interacts with Context Provider Registry (CPR) and Data Fusion Operator Repository (DFOR) and identifies which sensors and DFOs can be combined together to produce the remaining required context information. SCP identifies that \textbf{\textit{airStress}} can be calculated by using two other primary context information: \textbf{\textit{airTemperature}} and \textbf{\textit{airHumidity}}. Further, the numerical comparison operator and AND operator also required to derive \textbf{\textit{airStress}}. Once the primary and secondary context information acquisition mechanisms are identified, those detailed are sent to the Context and Semantic Discoverer Generator (CSDG).

\textbf{Step 6:} CSDG generates the Context and Semantic Context Discoverer (CSD) based on the details provided by PCP and SCP. CSD is equipped with all the necessary details that allow acquiring context data from appropriate sensors, applying necessary data fusion techniques and sending the data to the Data Dispatcher (DD).

\begin{figure}[h]
 \centering
\vspace{-0.23cm}	
 \includegraphics[scale=.73]{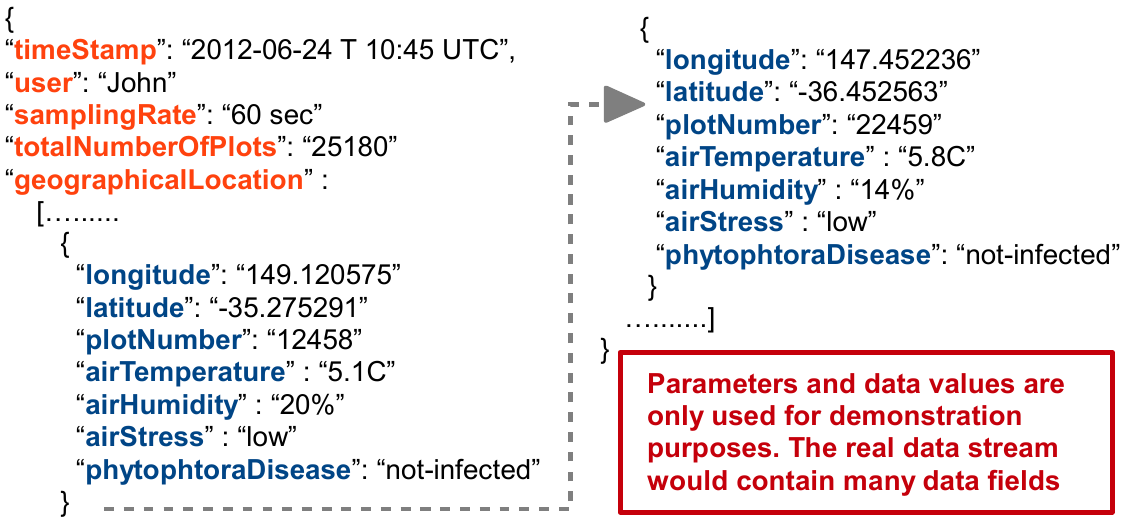}
\vspace{-0.23cm}	
 \caption{Sample data in JSON format which DD sends to John}
 \label{Sample data in JSON format which DD sends to John}	
\vspace{-0.33cm}	
\end{figure}

\textbf{Step 7:} DD retrieves user details from the P/S and transform the sensor data according to the user requested format. Finally, DD sends the data to the user. As depicted in Figure \ref{Sample data in JSON format which DD sends to John}, the final data stream combines number of data parameter such as \textbf{\textit{airTemperature, airHumidity, airStress, phytophtora-DiseaseStatus, timestamp, geographicalLocation}} and several other relevant context information that John can use to find answers to his problem. CA4IOT is not intended to provide direct answers the problems submitted by the users. Instead, it provides the user with necessary information that allows them to find the solution to their problem very easily.

\section{Conclusions and Future Work}
\label{sec:Conclusions and Future Work}

We identified through literature that there are significant amount of middleware solutions that exist for sensor data management related IoT, sensor networks, pervasive/ubiquitous computing, and context management fields. The problem is that they are strong in some aspect while weak in many other aspects. We also noticed that significant amount of solutions have overlapped and reinvented the wheel. Therefore, we decided to focus on one problem and propose a solution to tackle that problem where other IoT middleware projects can adopt our proposed architectural designs, models, and techniques to solve this particular problem within their own middleware solutions. 

This paper examined various aspects of context-aware IoT and presented the vision of CA4IOT. We presented an architecture that enhances the context aware capability of IoT middleware solutions and enables to build a sensing-as-a-service platform. The CA4IOT architecture has been proposed and designed with links to OpenIoT \cite{P377}, Phenonet \cite{P412} and SenseMA projects that CSIRO is involved in. We have implemented the components in sensor data acquisition layer (SDAL) and the details are presented in  \cite{ZMP002}. We will continue developing CA4IOT and project details, technical documents, and source code will be available in https://sourceforge.net/projects/CA4IOT/ in due course.

\def\IEEEbibitemsep{0pt plus .2pt}
\bibliographystyle{IEEEtran}
\bibliography{Bibliography}

\begin{thebibliography}{10}
\providecommand{\url}[1]{#1}
\csname url@samestyle\endcsname
\providecommand{\newblock}{\relax}
\providecommand{\bibinfo}[2]{#2}
\providecommand{\BIBentrySTDinterwordspacing}{\spaceskip=0pt\relax}
\providecommand{\BIBentryALTinterwordstretchfactor}{4}
\providecommand{\BIBentryALTinterwordspacing}{\spaceskip=\fontdimen2\font plus
\BIBentryALTinterwordstretchfactor\fontdimen3\font minus
  \fontdimen4\font\relax}
\providecommand{\BIBforeignlanguage}[2]{{%
\expandafter\ifx\csname l@#1\endcsname\relax
\typeout{** WARNING: IEEEtran.bst: No hyphenation pattern has been}%
\typeout{** loaded for the language `#1'. Using the pattern for}%
\typeout{** the default language instead.}%
\else
\language=\csname l@#1\endcsname
\fi
#2}}
\providecommand{\BIBdecl}{\relax}
\BIBdecl

\bibitem{P029}
H.~Sundmaeker, P.~Guillemin, P.~Friess, and S.~Woelffle, ``Vision and
  challenges for realising the internet of things,'' European Commission
  Information Society and Media, Tech. Rep., March 2010,
  http://www.internet-of-things-research.eu/pdf/IoT\_Clusterbook\_March\_2010.%
pdf [Accessed on: 2011-10-10].

\bibitem{P574}
{International Data Corporation (IDC) Corporate USA}, ``Worldwide smart
  connected device shipments,'' March 2012,
  \url{http://www.idc.com/getdoc.jsp?containerId=prUS23398412} [Accessed on:
  2012-08-01].

\bibitem{P009}
I.~Akyildiz, W.~Su, Y.~Sankarasubramaniam, and E.~Cayirci, ``A survey on sensor
  networks,'' \emph{Communications Magazine, IEEE}, vol.~40, no.~8, pp. 102 --
  114, aug 2002.

\bibitem{P035}
K.~Ellebek, ``A survey of context-aware middleware,'' in \emph{Proceedings of
  the 25th conference on IASTED International Multi-Conference: Software
  Engineering}.\hskip 1em plus 0.5em minus 0.4em\relax ACTA Press, 2007, pp.
  148--155.

\bibitem{P417}
M.~M. Molla and S.~I. Ahamed, ``A survey of middleware for sensor network and
  challenges,'' in \emph{Proceedings of the 2006 International Conference
  Workshops on Parallel Processing}, ser. ICPPW '06.\hskip 1em plus 0.5em minus
  0.4em\relax Washington, DC, USA: IEEE Computer Society, 2006, pp. 223--228.

\bibitem{P118}
S.~Bandyopadhyay, M.~Sengupta, S.~Maiti, and S.~Dutta, ``Role of middleware for
  internet of things: A study,'' \emph{International Journal of Computer
  Science and Engineering Survey}, vol.~2, pp. 94--105, 2011.

\bibitem{P291}
P.~Bellavista, A.~Corradi, M.~Fanelli, and L.~Foschini, ``A survey of context
  data distribution for mobile ubiquitous systems,'' \emph{ACM Computing
  Surveys}, p.~49, 2013.

\bibitem{P504}
J.~Manyika, M.~Chui, B.~Brown, J.~Bughin, R.~Dobbs, C.~Roxburgh, and A.~H.
  Byers, ``Big data: The next frontier for innovation, competition, and
  productivity,'' McKinsey Global Institute, Tech. Rep., May 2011,
  \url{http://www.mckinsey.com/Insights/MGI/Research/Technology_and_Innovation%
/Big_data_The_next_frontier_for_innovation} [Accessed on: 2012-06-08].

\bibitem{P022}
K.~Aberer, M.~Hauswirth, and A.~Salehi, ``Infrastructure for data processing in
  large-scale interconnected sensor networks,'' in \emph{International
  Conference on Mobile Data Management}, 2007, pp. 198--205.

\bibitem{P227}
{GSN Team}, ``Global sensor networks project,'' 2011,
  \url{http://sourceforge.net/apps/trac/gsn/} [Accessed on: 2011-12-16].

\bibitem{P498}
S.~Patidar, D.~Rane, and P.~Jain, ``A survey paper on cloud computing,'' in
  \emph{Advanced Computing Communication Technologies (ACCT), 2012 Second
  International Conference on}, jan. 2012, pp. 394 --398.

\bibitem{ZMP001}
C.~Perera, A.~Zaslavsky, P.~Christen, A.~Salehi, and D.~Georgakopoulos,
  ``Capturing sensor data from mobile phones using global sensor network
  middleware,'' in \emph{IEEE International Workshop on Internet-of-Things
  Communications and Networking 2012 (PIMRC 2012-Workshop-IoT-CN12)}, Sydney,
  Australia, September 2012.

\bibitem{ZMP002}
------, ``Connecting mobile things to global sensor network middleware using
  system-generated wrappers,'' in \emph{International ACM Workshop on Data
  Engineering for Wireless and Mobile Access 2012 (ACM SIGMOD/PODS
  2012-Workshop-MobiDE)}, Scottsdale, Arizona, USA, May 2012.

\bibitem{P257}
{NIST}, ``Introduction to ieee 1451,'' 2011,
  \url{http://www.nist.gov/el/isd/ieee/1451intro.cfm} [Accessed on:
  2012.03.01].

\bibitem{P083}
M.~Botts, G.~Percivall, C.~Reed, and J.~Davidson, ``Ogc sensor web enablement:
  Overview and high level architecture,'' in \emph{Geosensor Networks Lecture
  Notes in Computer Science}, S.~Nittel, A.~Labrinidis, and A.~Stefanidis,
  Eds.\hskip 1em plus 0.5em minus 0.4em\relax Berlin, Heidelberg:
  Springer-Verlag, 2008, vol. 4540, pp. 175--190.

\bibitem{P382}
{Oracle Corporation}, ``Sun spot world: Welcome to the internet of things,''
  2012, http://www.sunspotworld.com/ [Accessed on: 2012-04-10].

\bibitem{P185}
M.~Perttunen, J.~Riekki, and O.~Lassila, ``Context representation and reasoning
  in pervasive computing: a review,'' \emph{International Journal of Multimedia
  and Ubiquitous Engineering}, vol.~4, no.~4, pp. 1--28, 2009.

\bibitem{P452}
A.~Baggio, ``Wireless sensor networks in precision agriculture,'' Delft
  University of Technology – The Netherlands, Tech. Rep., 2009,
  \url{http://www.sics.se/realwsn05/papers/baggio05wireless.pdf} [Accessed on:
  2012-05-10].

\bibitem{P377}
{OpenIoT Consortium}, ``Open source solution for the internet of things into
  the cloud,'' January 2012, http://www.openiot.eu/ [Accessed on: 2012-04-08].

\bibitem{P412}
{CSIRO}, ``Phenonet: Distributed sensor network for phenomics supported,''
  2011, http://phenonet.com/ [Accessed on: 2012-04-20].

\end{thebibliography}
\footnotesize

\end{document}